\documentclass[submission,copyright,creativecommons]{eptcs}
\providecommand{\event}{QPL 2014} % Name of the event you are submitting to
\usepackage{epsfig,amssymb,amsmath,mathrsfs} 
 \newtheorem{proposition}{Proposition}
 \newtheorem{theorem}{Theorem}
 
\newtheorem{ax}{Axiom}

\def\map #1{{\mathcal #1}}

 \def\>{\rangle}
\def\<{\langle} \def\trnsfrm#1{\mathcal #1}
 \def\rA{{\rm A}}\def\rB{{\rm
    B}}\def\rC{{\rm C}}\def\rD{{\rm D}} \def\rE{{\rm E}} \def\rF{{\rm
    F}} \def\rG{{\rm G}} 
    \def\rH{{\rm H}}

    \def\rI{{\rm I}} \def\rR{{\rm R}} 
     \def\rX{{\rm X}}
     \def\rY{{\rm Y}}

\def\tU{\trnsfrm U}
 
\def\tM{\trnsfrm M}
\def\tN{\trnsfrm N} 
\def\tP{\trnsfrm P}   
 \def\Eff{{\mathsf {Eff}}}
 \def\Span{\mathsf{Span}}
\def\St{{\mathsf {St}}} 
\def\Transf{{\mathsf {Transf}}}
  
  \def\Tr{{\rm Tr}}
 \def\Reals{{\mathbb R}}

\def\Pur{{\sf Pur}}

\def\Op{{\sf Op}}
\def\Rev{{\sf Rev}}
\def\Det{{\sf Det}}
\def\Sys{{\sf Sys}}
\def\Tests{{\sf Tests}}

\def\Outcomes{{\sf Outcomes}}

\def\Prob{{\sf Prob}}

\usepackage[matrix,frame,arrow]{xy}
\newcommand{\qw}[1][-1]{\ar @{-} [0,#1]}
\newcommand{\gate}[1]{*{\xy *+<.6em>{#1};p\save+LU;+RU **\dir{-}\restore\save+RU;+RD **\dir{-}\restore\save+RD;+LD **\dir{-}\restore\POS+LD;+LU **\dir{-}\endxy} \qw}
\newcommand{\measureD}[1]{*{\xy*+=+<.5em>{\vphantom{\rule{0em}{.1em}#1}}*\cir{r_l};p\save*!R{#1} \restore\save+UC;+UC-<.5em,0em>*!R{\hphantom{#1}}+L **\dir{-} \restore\save+DC;+DC-<.5em,0em>*!R{\hphantom{#1}}+L **\dir{-} \restore\POS+UC-<.5em,0em>*!R{\hphantom{#1}}+L;+DC-<.5em,0em>*!R{\hphantom{#1}}+L **\dir{-} \endxy} \qw}
\newcommand{\multimeasureD}[2]{*+<1em,.9em>{\hphantom{#2}}\save[0,0].[#1,0];p\save !C *{#2},p+LU+<0em,0em>;+RU+<-.8em,0em> **\dir{-}\restore\save +LD;+LU **\dir{-}\restore\save +LD;+RD-<.8em,0em> **\dir{-} \restore\save +RD+<0em,.8em>;+RU-<0em,.8em> **\dir{-} \restore \POS !UR*!UR{\cir<.9em>{r_d}};!DR*!DR{\cir<.9em>{d_l}}\restore \qw}
\newcommand{\multigate}[2]{*+<1em,.9em>{\hphantom{#2}} \qw \POS[0,0].[#1,0];p !C *{#2},p \save+LU;+RU **\dir{-}\restore\save+RU;+RD **\dir{-}\restore\save+RD;+LD **\dir{-}\restore\save+LD;+LU **\dir{-}\restore}
\newcommand{\ghost}[1]{*+<1em,.9em>{\hphantom{#1}} \qw}
\newcommand{\Qcircuit}[1][0em]{\xymatrix @*=<#1>} % @*[o]
\newcommand{\pureghost}[1]{*+<1em,.9em>{\hphantom{#1}}}
\newcommand{\multiprepareC}[2]{*+<1em,.9em>{\hphantom{#2}}\save[0,0].[#1,0];p\save !C
  *{#2},p+RU+<0em,0em>;+LU+<+.8em,0em> **\dir{-}\restore\save +RD;+RU **\dir{-}\restore\save
  +RD;+LD+<.8em,0em> **\dir{-} \restore\save +LD+<0em,.8em>;+LU-<0em,.8em> **\dir{-} \restore \POS
  !UL*!UL{\cir<.9em>{u_r}};!DL*!DL{\cir<.9em>{l_u}}\restore}
\newcommand{\prepareC}[1]{*{\xy*+=+<.5em>{\vphantom{#1\rule{0em}{.1em}}}*\cir{l^r};p\save*!L{#1} \restore\save+UC;+UC+<.5em,0em>*!L{\hphantom{#1}}+R **\dir{-} \restore\save+DC;+DC+<.5em,0em>*!L{\hphantom{#1}}+R **\dir{-} \restore\POS+UC+<.5em,0em>*!L{\hphantom{#1}}+R;+DC+<.5em,0em>*!L{\hphantom{#1}}+R **\dir{-} \endxy}}
\newcommand{\poloFantasmaCn}[1]{{{}^{#1}_{\phantom{#1}}}}

\title{
Dilation of states and processes in operational-probabilistic theories}
\author{Giulio Chiribella
\institute{Center for Quantum Information, Institute for Interdisciplinary Information Sciences, \\
Tsinghua University, Beijing, 100084, China}
\email{gchiribella@mail.tsinghua.edu.cn}}
\def\titlerunning{Dilation of states and processes in operational-probabilistic theories}
\def\authorrunning{G. Chiribella}
\begin{document}
\maketitle

\begin{abstract}
This paper  provides a concise summary of the framework of operational-probabilistic theories, aimed at  emphasizing the interaction between  category-theoretic and probabilistic structures.  Within this framework,  we review an operational version of the GNS construction,  expressed by the so-called  purification principle \cite{chiribella10}, which under mild hypotheses leads to an operational version of Stinespring's theorem.    
\end{abstract}
\bigskip 

\section{Introduction} 

Few theories in physics have been as successful and as surprising  as Quantum Theory.    Both successes and surprises come from a simple mathematical framework of virtually universal applicability, which blends physics and information theory in peculiar and often puzzling way. 
At the heart of this framework  is a set of mathematical theorems, known as \emph{dilation theorems} \cite{paulsenbook}, which allow one to reduce all possible states, evolutions, and measurements allowed by quantum mechanics to some privileged subsets.    
 Specifically,  
\begin{enumerate}
\item mixed states are reduced to pure states (by the GNS construction \cite{GN,S}, familiar to the quantum information community as \emph{purification}) 
\item general evolutions are reduced to reversible evolutions (by Stinespring's theorem \cite{stine} )
\item  general measurements are reduced to sharp measurements (by Naimark's \cite{naimark} and Ozawa's \cite{ozawa} theorems).  
\end{enumerate}  
For finite dimensional quantum systems and for the simplest examples of infinite dimensional systems (type 1 von Neumann algebras), the reductions 1-3 are  achieved by introducing  an auxiliary system   (the \emph{environment}), which is eventually discarded.  This fact lends itself to an operational interpretation: The ignorance about the preparation of a system, the irreversibility of an evolution, and the unsharpness of a measurement can always  be explained as resulting from the lack of control over some degree of freedom  in the surrounding environment.    

Dilation theorems are usually regarded as a consequence of the mathematical framework of Quantum Theory.  They are heavily employed as  technicals tools by researchers in quantum information theory, to the extent that one can hardly  find results that do not invoke any of them, at least  in an   implicit way.   The dilation approach is so common in quantum information that it earned itself a nickname---the ``Church of the Larger Hilbert Space" \footnote{The  expression is due to John Smolin, see e.g. the lecture notes \cite{bennett}.}.   
However, the operational content of the dilation  theorems is independent of the quantum framework:   Even forgetting  about Hilbert spaces and operator algebras, one can still express the notions of pure/mixed state, reversible/irreversible evolution, and sharp/unsharp measurement in a general framework of operational-probabilistic theories  \cite{hardy2001,barrett,mauro2,nobroad,teleportation,chiribella10,mauro1,chiribella11,barnum2011,hardybig,hardy2013}. In this broader framework the dilation of states, evolutions and measurements can be promoted  to the rank of \emph{axioms}, from which (a number of features of) the theory  is derived    (\cite{chiribella10})  \cite{chiribella11}. 
 There are at least three good reasons to follow this route.  First,  given the amount of results that invoke dilation theorems in quantum  information processing, turning these theorems into axioms seems to be a convenient way to restructure the landscape of quantum information and to facilitate the discovery of new protocols. 
Second, the dilation approach sheds light on the old question ``Why the quantum?", the question of finding a set of well-motivated axioms that single out quantum theory among all possible theories.  This is the path followed by Ref. \cite{chiribella11} where the finite-dimensional Hilbert space framework has been reconstructed from the purification of mixed states---the so-called \emph{Purification Principle}.  In the light of this result, Quantum Theory appears as the golden standard of theory where information-theoretic notions admit a description  in terms of pure states and fundamentally reversible interactions \cite{chiribella12}.   Third, the approach of abstracting dilation theorems from the Hilbert space framework yielded deep insights  in category theory, leading  to the formulation of Selinger's  CPM construction \cite{CPM} and to its axiomatization in terms of interaction with the environment \cite{coecke2008axiomatic,coecke2010environment}.

In the following we review the framework of operational-probabilistic theories and  the basic results about  the  Purification Principle,    providing  operational versions of  the GNS construction and of Stinespring's theorem in the finite-dimensional setting.    
The goal of this presentation is to provide a concise and mathematically rigorous summary of the state of the art.   Overall, its contribution does not consist in new results, but rather in the systematization of the existing ones, which are put here in a more compact form.    Since this is meant to be a summary of results, we will not provide proofs, most of which can anyways be recovered from the original work \cite{chiribella10}.  
%Also, we will not treat here the operational version of  Naimark-type theorems.  For them and for one possible definition of sharp measurements in the operational context we refer the interested reader to Ref.  \cite{sharp}.     

\section{The framework of operational-probabilistic theories}  
In this section we review the framework of operational-probabilistic theories \cite{chiribella10}. The framework consists of  two distinct conceptual ingredients: an operational structure, describing circuits that produce outcomes, and a probabilistic structure, which assigns probabilities to the outcomes in a consistent way.   
The operational structure is closely related to  and partly inspired by  the area of categorical quantum mechanics  \cite{abramsky2004,coecke2006,abramsky2008}. There are nevertheless a few relevant  differences in the way classical outcomes are treated, which turn out to be important when introducing the probabilistic structure.

\subsection {Operational structure}
The operational structure  summarizes  all the possible circuits that can be constructed in a given physical theory.  In general, the circuits can consist of non-deterministic gates, where  evolution of the input   branches      into a number of alternative processes  $\{\map M_x\}_{x\in\rX}$ labelled by a finite index set $\rX$.    

% In addition to the outcome set $\rX$, a gate has an input system $\rA$ and an output system $\rB$.  

%Graphically, we represent the non-deterministic gate   $\{\map M_x\}_{x\in\rX}$ as
%\begin{align*}
 %\begin{aligned} &\Qcircuit @C=1em @R=.7em @! R {
  %  &\poloFantasmaCn{\rA}\qw&\gate{\{\tM_x\}_{x\in\rX}}&\poloFantasmaCn{\rB}\qw& }  
   % \end{aligned} \quad . 
    %\end{align*}
    \medskip 
    
\subsubsection{The category of transformations}    For a given non-deterministic gate  $\{\map M_x\}_{x\in\rX}$,  each process $\map M_x$  is regarded as a   physical transformation, transforming an input system $\rA$ into an output system $\rB$.  The set of all physical transformations is required to be a  (strict) symmetric monoidal category  (SMC) \cite{awodey2006}, denoted by $\Transf$, and the set of all transformations of type $\rA\to \rB$ is denoted as $\Transf(\rA\to \rB)$.    The transformations of type $\rI$ to $\rI$ are also refereed to as \emph{scalars.}   The identity morphism on system $\rA$  is called the  \emph{identity transformation} and is denoted by $\map I_\rA$.     The identity on the tensor unit $\rI$ will be denoted by $\sf 1$.  
      The sequential and parallel composition of transformations are represented in the graphical language of  SMCs  \cite{selinger2011survey,coecke2010quantum}, which consists of  circuits like
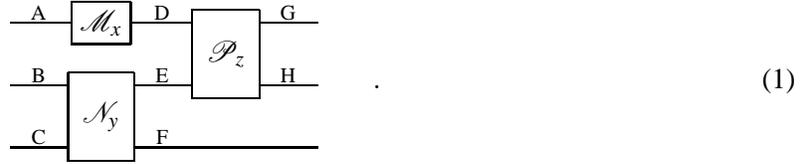
\begin{align}\label{example2}
 \begin{aligned} &\Qcircuit @C=1em @R=.7em @! R {
    &\poloFantasmaCn{\rA}\qw&\gate{\tM_x  }&\poloFantasmaCn{\rD}\qw& \multigate{1}{\tP_z}   &   \poloFantasmaCn{  \rG} \qw &\qw\\
     &\poloFantasmaCn{\rB}\qw&\multigate{1}{\tN_y}&\poloFantasmaCn{\rE}\qw&   \ghost{\tP_z}     &   \poloFantasmaCn{\rH} \qw &\qw\\
     &\poloFantasmaCn{\rC}\qw&\ghost{\tN_y }&  \poloFantasmaCn{\rF}  \qw& \qw  &  \qw  &\qw  }
 \end{aligned}   \quad .
\end{align}

\subsubsection{The category of tests}  A non-deterministic gate of type $\rA\to \rB$ is a indexed list of transformations in $\{\map M_x\}_{x\in\rX}  \subset \Transf(\rA\to \rB)$.    The index set $\rX$   is called the \emph{outcome space} and represents the possible outcomes that can occur when the gate is used.  We denote by $\Outcomes$  the set of all possible outcome spaces appearing in the theory. For reasons that will be immediately clear, we require that 
\begin{enumerate}
\item  $\Outcomes$ is closed under Cartesian product
\item $\Outcomes$  contains the singleton set $\{\epsilon\}$, where $\epsilon$ is the empty word, so that $\{\epsilon\}  \times  \rX  =   \rX\times \{\epsilon\}  =  \rX$ for every  $\rX\in\Outcomes$.   
\end{enumerate}

Following the terminology of Ref. \cite{chiribella10}, we refer to non-deterministic gates as \emph{tests} and we denote the tests of type $\rA\to \rB$ and with outcome set $\rX$  as  $\Tests (\rA\to \rB,  \rX)$.   The set of all tests of type $\rA  \to \rB$ will be denoted as  
$$   \Tests (\rA\to \rB)  :  =  \bigcup_{\rX\in\Outcomes}    \Tests (\rA\to \rB,  \rX) \, .    $$ 

The sequential composition in $\Transf$  induces a sequential composition of tests, defined as     
 \begin{align}\label{seq}
 \{\tN_y\}_{y\in\rY}   \circ   \{\tM_x\}_{x\in\rX}  : =  \{  \map N_y \circ \map M_x\}_{(x,y)\in  \rX\times \rY} \, .  
 \end{align}
By this definition, the sequential composition of a test   in  $\Tests  (\rA\to \rB,\rX)$ with a test in $\Tests (\rB\to \rC,  \rY)$ yields a test in  $  \Tests  (\rA\to \rC,  \rX\times \rY)$.  
The \emph{identity test on system $\rA$} is the test   $\{\map I_\rA\}$  with singleton outcome set $\{\epsilon\}$. 

The parallel composition in $\Transf $ induces a parallel composition of tests, defined as        
\begin{align}\label{par}
  \{\tM_x\}_{x\in\rX}  \otimes  \{\tN_y\}_{y\in\rY}     : =  \{   \map M_x  \otimes \map N_y   \}_{(x,y)\in  \rX\times \rY} \, .  
 \end{align}
By this definition,  the parallel composition of  a test   in  $\Tests  (\rA\to \rA',\rX)$ with a test in $\Tests (\rB\to \rB',  \rY)$ yields a test in  $  \Tests  (\rA  \otimes \rB\to \rA'\otimes \rB',  \rX\times \rY)$.  

It is immediate to verify that the collection of all tests forms a SMC, where the systems are the same systems of $\Transf$,  the morphisms are indexed lists of morphisms in $\Transf$, and the operations of sequential and parallel composition are defined as in Eqs. (\ref{seq}) and (\ref{par}).   We denote this category by $\Tests$.

Summing up, the operational structure of a theory is described by a triple 
$$\Op  :=    ( \Transf,\Outcomes,  \Tests)  \, ,$$which specifies which transformations are physically possible, which outcomes herald these transformations, and which tests can be performed.
% and which circuits can be build from these tests. 
%Occasionally, we use the graphical language also for the category of tests, representing circuits like
%\begin{align}\label{example}
 %\begin{aligned} &\Qcircuit @C=1em @R=.7em @! R {
 %   &\poloFantasmaCn{\rA}\qw&\gate{\{\tM_x\}_{x\in\rX}}&\poloFantasmaCn{\rD}\qw& \multigate{1}{\{\tP_z\}_{z\in\rZ}}   &   \poloFantasmaCn{  \rG} \qw &\qw\\
 %    &\poloFantasmaCn{\rB}\qw&\multigate{1}{\{\tN_y\}_{y\in\rY}}&\poloFantasmaCn{\rE}\qw&   \ghost{\{\tP_z\}_{z\in\rZ}}     &   \poloFantasmaCn{\rH} \qw &\qw\\
  %   &\poloFantasmaCn{\rC}\qw&\ghost{\{\tN_y\}_{y\in\rY} }&  \poloFantasmaCn{\rF}  \qw& \qw  &  \qw  &\qw  }
 %\end{aligned}   \, .
%\end{align}
%A circuit of tests represents an array of non-deterministic gates, which can give raise to many different combinations of outcomes.   For every individual combination one will have an array  of physical transformations, as in Eq. (\ref{example2}).  

\subsection{Probabilistic structure} 
    
 The probabilistic structure of a theory consists of a rule to assign probabilities to the outcomes that are generated by non-deterministic circuits.    Precisely, the rule is given by a function  $\Prob $ which maps the scalars   $\Transf  (\rI\to \rI) $ into real numbers  in the interval    $[0,1]\subset \Reals$ and satisfies  the following properties 
\begin{enumerate}
\item \emph{Consistency:}  for   every outcome space $\rX\in\Outcomes$ and for every  test $\{s_x\}_{x\in\rX}  \in \Tests(\rI\to \rI,\rX)$    the function $\Prob \circ  s$ is a probability distribution,~i.~e.    $\Prob(s_x)  \ge 0$ and $\sum_{x\in\rX}    \Prob (s_x)  =  1$. 
\item \emph{Independence:}     $ {\sf Prob}  ( s    \otimes t)   =   {\sf Prob}  (s) \,    {\sf Prob}(t)$ for every pair of scalars  
$s,t\in\Transf(\rI\to\rI)$. 
\end{enumerate}
Here the consistency property guarantees that we can interpret ${\sf Prob} (s_x) $ as the probability of the outcome $x\in\rX$.    The independence property guarantees that experiments that involve only independent tests on two systems give rise to uncorrelated outcomes.  As observed by Hardy \cite{hardy2013,hardybig}, independence is equivalent to the requirement that  one can assign probabilities to a closed circuit in a way that is independent of the context.  
Note that the map $\Prob$ does not need to  be onto: for example, in a deterministic theory the range of the map $\Prob$ are only the values $0$ and $1$.

The probabilistic structure endows the scalars in $\Transf$ with a structure of \emph{test space}  \cite{wilce2009test,wilce2000test}. Mathematically, a test space is a hypergraph equipped with function that assigns  probabilities to vertices, under the condition that the sum of the probabilities is equal to 1 for every hyperedge.       Note that the requirement of independence in the definition above adds extra structure to the test space, forcing  a homomorphism between the monoid of scalars and the monoid of probabilities.      Test spaces with additional structure have been recently considered in the characterization of contextuality \cite{cabello2010,cabello2014}  and non-locality \cite{acin2012}.

Combining the ingredients given so far, we can give the formal definition of an operational-probabilistic theory as  a couple $\Theta  :  =  (  \Op,\Prob) $, or equivalently, as a quadruple $\Theta  =  (  \Transf,\Outcomes,\Tests,\Prob)$.

\subsection{Quotient theories}

The interaction between the operational structure and the  probabilistic structure has major consequences.   Eventually, it allows one to represent  the category of physical transformations as a category of positive maps on ordered vector spaces.  The steps that lead to this representation are  summarized in the following.

\subsubsection{The operation of quotient}  The probabilistic structure brings with itself a natural notion of equivalent transformations.  To spell out this notion clearly, it is convenient to introduce some notation:    a transformation   $\rho$ of type  $\rI\to \rA$ will be called a \emph{state of system $\rA$} and will be represented as 
\begin{align}
\begin{aligned}  
\Qcircuit @C=.5em @R=0em @!R {     &\prepareC {\rho}  & \qw \poloFantasmaCn{\rA}    & \qw} 
\end{aligned}     
:= 
\begin{aligned}  
\Qcircuit @C=.5em @R=0em @!R {     & \qw \poloFantasmaCn{\rI}  &\gate{\rho}   &   \qw \poloFantasmaCn{\rA}   & \qw} 
\end{aligned}    \quad .
\end{align}     
The set of all states of system $\rA$ will be denoted as $\St(\rA)$.    
A transformation $a$ of type $\rA\to \rI$ will be called an \emph{effect on system $\rA$} and will be represented as 
\begin{align}
\begin{aligned}  
\Qcircuit @C=.5em @R=0em @!R {  & \qw \poloFantasmaCn{\rA}    & \measureD a} 
\end{aligned}     
:= 
\begin{aligned}  
\Qcircuit @C=.5em @R=0em @!R {     & \qw \poloFantasmaCn{\rA}  &\gate{a}   &   \qw \poloFantasmaCn{\rI}   & \qw} 
\end{aligned}    \quad .
\end{align}       
 The set of all effects on system $\rA$ will be denoted as $\Eff(\rA)$.      Two transformations of type $\rA\to \rB$, say $\map M$ and $\map M'$, are called \emph{equivalent} iff  
\begin{align}\label{equitransf}
 {\sf Prob}  \left(  
 \begin{aligned}
 \Qcircuit @C=1em @R=.7em @! R { 
    \multiprepareC{1}{\rho}&\poloFantasmaCn{\rA}\qw&\gate{\tM}   &   \poloFantasmaCn{\rB}  \qw  &\multimeasureD{1}{E}\\
     \pureghost{\rho}&\poloFantasmaCn{\rR}\qw&\qw & \qw&\ghost{E}}  
     \end{aligned}  
    \right)  
    =
     {\sf Prob} \left(  \begin{aligned}
     \Qcircuit @C=1em @R=.7em @! R { 
    \multiprepareC{1}{\rho}&\poloFantasmaCn{\rA}\qw&\gate{\tM'}     &  \poloFantasmaCn{\rB}  \qw  &\multimeasureD{1}{E}\\
     \pureghost{\rho}&\poloFantasmaCn{\rR}\qw&\qw &\qw& \ghost{E}}   
    \end{aligned}  
    \right)  
    \end{align}
for every system $\rR\in\Sys$, every state $\rho\in\St(\rA\otimes \rR)$, and every effect $E\in\Eff(\rB\otimes \rR)$.  
 
Note that the use of the ``reference system" $\rR$ is essential, unless  the theory enjoys a property known as ``local tomography". \cite{hardy2001,barrett,mauro2}.  When local tomography is not satisfied,    checking the validity of Eq. (\ref{equitransf}) only for $\rR  = \rI$ may not be sufficient to guarantee its validity for arbitrary $\rR$.  
There are only two cases where the system $\rR$ is not needed by default:  the case of states (transformations of type $\rI\to \rA$) and the case of effects (transformations of type $\rA\to \rI$).   For example,  for two states  $\beta,\beta'\in\St(\rB)$ Eq. (\ref{equitransf}) reads  
\begin{align*}
 {\sf Prob}  
 \left(  
 \begin{aligned}
 \Qcircuit @C=1em @R=.7em @! R { 
      \prepareC{\beta }   &   \poloFantasmaCn{\rB}  \qw  &\multimeasureD{1}{E}\\
      \prepareC{\rho}&\poloFantasmaCn{\rR}  \qw & \ghost{E}}  
     \end{aligned}  
    \right)  
    =
     {\sf Prob}
      \left(  \begin{aligned}
     \Qcircuit @C=1em @R=.7em @! R { 
    &\prepareC{\beta}     &  \poloFantasmaCn{\rB}  \qw  &\multimeasureD{1}{E}   \\
       &\prepareC{\rho}&\poloFantasmaCn{\rR}\qw& \ghost{E}      }   
    \end{aligned}  
    \right)     \qquad \qquad    \forall  \rho\in \St(\rR)    \,  ,  \forall  E\in \St(\rB\otimes \rR)  \, ,
    \end{align*}  
    which is equivalent to the condition   
\begin{align*}
 {\sf Prob}  
 \left(  
 \begin{aligned}
 \Qcircuit @C=1em @R=.7em @! R { 
      \prepareC{\beta }   &   \poloFantasmaCn{\rB}  \qw  &\measureD{b} }  
     \end{aligned}  
    \right)  
    =
     {\sf Prob}
      \left(  \begin{aligned}
     \Qcircuit @C=1em @R=.7em @! R { 
    \prepareC{\beta}     &  \poloFantasmaCn{\rB}  \qw  &\measureD{b} }   
    \end{aligned}  
    \right)     \qquad \qquad       \forall  b\in \St(\rB)  \, .
    \end{align*}

We denote by $[\map M]$ the equivalence class of the transformation $\map M$ and by $\left[\Transf(\rA\to \rB)\right]$ the set of all equivalence classes of transformations of type $\rA\to \rB$.  It is easy to see that the  equivalence classes of transformations form an SMC, denoted by       $[\Transf]$, and that the  operation of quotient  is a forgetful (strong symmetric monoidal) functor from $\Transf$ to    $[\Transf]$. 
 Furthermore, the equivalence classes of transformations induce equivalence classes of tests,  defined as $$\left[  \{\map M\}_{x \in  \rX}    \right  ]  :=    \left\{  \,[  \map M_x] \, \right\}_{x\in \rX} .$$    
 Also in this case the operation of quotient is a forgetful (strong symmetric monoidal) functor, from the SMC $\Tests$ to the SMC $[\Tests]$ consisting of equivalence classes of tests.    In summary the operational structure $\Op  =  (\Transf,\Outcomes,\Tests)$  can be mapped functorially into a new operational structure $$[\Op]  :=  (  [\Transf],  \Outcomes, [\Tests]) \, .$$  Defining the probabilistic structure $[\Prob]$ by the relation
$$ [ \Prob]  (  [ s]   )   :  =  \Prob (s)  \qquad \forall s\in\Transf(\rI\to \rI) \, ,$$ 
it is easy to check that the couple $[\Theta]  =   (  [\Op], [\Prob])$ is an operational-probabilistic theory, which we call the \emph{quotient theory}.

\subsubsection{Axiomatic characterization of quotient theories}

From now on, we will \emph{only} consider quotient theories: given an operational-probabilistic theory  $\Theta  =  (\Transf,  \Outcomes, \Tests,\Prob)$, we will assume that the operation of quotient has been already made.   This is equivalent to requiring  the following \emph{separation axiom}:       
\begin{ax}[Separation]\label{ax:separation}
If  $\map M$ and $\map M'$ are such that 
\begin{equation*}
\Prob\left( \begin{aligned}  \Qcircuit @C=.5em @R=0em @!R { 
& \multiprepareC{2}{\rho}    & \qw \poloFantasmaCn{\rA}  &  \gate{\map M}  &  \qw \poloFantasmaCn{\rB}  &\multimeasureD{2}{E} \\
& \pureghost{\rho}   &&&&\pureghost{E}\\
 & \pureghost{\rho}    & \qw \poloFantasmaCn{\rR}  &  \qw &\qw &\ghost{E}} 
\end{aligned}  \quad  \right)  
=
\Prob\left(\begin{aligned}  \Qcircuit @C=.5em @R=0em @!R { & \multiprepareC{2}{\rho}    & \qw \poloFantasmaCn{\rA}  &    \gate{\map M'}  &  \qw \poloFantasmaCn{\rB}  &\multimeasureD{2}{E}  \\
& \pureghost{\rho}   &&&&\pureghost{E}\\
 & \pureghost{\rho}    & \qw \poloFantasmaCn{\rR}  &  \qw&\qw&\ghost{E} } 
\end{aligned} \quad \right)      \qquad 
\begin{array}{l}
\forall \rR\in\Sys \\
\forall \rho\in\St(\rA\otimes \rR)  \\
 \forall  E\in\Eff(\rB\otimes \rR)
\end{array}  
 \end{equation*}
 then $\map M  =\map M'$.  
\end{ax}

%Thanks to the Separation Axiom, one can  identify scalars with probabilities. 
%As a result, one can eliminate $\Prob$ from the equations, writing~e.~g.
%\begin{equation}\label{easierlife}
%\begin{aligned}\Qcircuit @C=.5em @R=0em @!R { & \multiprepareC{2}{\rho}    & \qw \poloFantasmaCn{\rA}  &    \gate{\map M}  &  \qw \poloFantasmaCn{\rB}  &\multimeasureD{2}{E}  \\
%& \pureghost{\rho}   &&&&\pureghost{E}\\
% & \pureghost{\rho}    & \qw \poloFantasmaCn{\rR}  &  \qw&\qw&\ghost{E} } 
%\end{aligned}   \qquad{\rm in~place~of} \qquad
%\Prob \left( \begin{aligned}  \Qcircuit @C=.5em @R=0em @!R { 
%& \multiprepareC{2}{\rho}    & \qw \poloFantasmaCn{\rA}  &  \gate{\map M}  &  \qw \poloFantasmaCn{\rB}  &\multimeasureD{2}{E} \\
%& \pureghost{\rho}   &&&&\pureghost{E}\\
% & \pureghost{\rho}    & \qw \poloFantasmaCn{\rR}  &  \qw &\qw &\ghost{E}} 
%\end{aligned}  \quad  \right)    \quad .
 %\end{equation}

\medskip

\subsubsection{Embedding into ordered vector spaces}    Quotient theories have  one major feature, highlighted by the following 
\begin{theorem}\label{theo:emb}
Let $ \Transf$ be the category of transformations in a quotient theory.  Then, there exists a  symmetric monoidal functor  $F$ from  $\Transf$ to an SMC   where the objects are partially order real vector spaces and the morphisms are positive linear maps.  
The functor $F$ has the following properties: 
\begin{enumerate}
\item $F(\rI)  =  \Reals$
\item  For every system $\rA  \in  |\Transf|$,   one has
\begin{align}\label{FA}
    \Span_{\Reals}    \left\{ F  (  \alpha)   \, ,~  \alpha \in \St (\rA)   \right\}  \simeq  F(\rA)  \, , 
    \end{align}
    \begin{align}\label{FA*}
      \Span_{\Reals}    \left\{ F  (  a)   ~,~  a \in \Eff (\rA)   \right\}   \simeq F(\rA)^*       \, , 
    \end{align}
     \begin{align}\label{probF}
F(  a)  \circ F(\alpha)   =       \Prob  (a\circ\alpha)     \qquad \forall \alpha\in\St(\rA)\, ,\forall  a\in\Eff  (\rA) \, .   
     \end{align}
%\item      for every system $\rA  \in  |\Transf|$,   the positive cone in $F(\rA)$ is generated by the vectors of the form  $F(\alpha)$ for $\alpha\in\St(\rA)$
\item   For every pair of systems $\rA,\rB\in  |\Transf|$ and every pair  of transformations $\map M, \map M'  \in  \Transf(\rA\to \rB)$, one has the separation property
\begin{align}\
  F  (      \map M'\otimes \map I_\rR)      =           F  (      \map M\otimes \map I_\rR)     \quad   &\forall  \rR\in  |\Transf|  
   \quad \Longrightarrow \quad    \map M'  =  \map M  \, . %\\
   %\nonumber     &     \forall  \Psi\in\St(\rA\otimes \rR)  \\
     %  &   \forall  E  \in  \Eff (\rB\otimes \rR)  
     \label{separation}
\end{align}
  
\end{enumerate}
\end{theorem}
The correspondence carries over to the tests in  $\Tests$, which can be mapped functorially into an SMC  of partially ordered vector spaces and indexed lists of positive maps.   

An important point needs to be made  here: in general the functor $F$ is \emph{neither faithful nor strong}.      It is not faithful, because if  the local tomography property does not hold, one can have  $F(\map M')  =   F(\map M)$ even if $\map M'  \not  =  \map M$.    This is not in contradiction with the separation property of Eq. (\ref{separation}), which only implies that there exists some system $\rR\in |\Transf|$ such that  
$F  (\map M'  \otimes \map I_\rR)  \not  =  F(\map M\otimes \map I_\rR) \, .$
The separation property also implies that, if $F$ is not faithful,   it cannot be strong,~i.~e.~one has
\begin{align}\label{strong} 
F(\map M\otimes \map N)   \not \simeq   F(\map M)  \otimes F(\map N) \, .
\end{align}  

To some extent, this is a bug that needs to be fixed, because it would be very convenient to identify  physical transformations with positive maps.  
The fix is provided by the separation property of Eq. (\ref{separation}),  which guarantees that   the transformation $\map M$  is in one-to-one correspondence with the indexed set   $  \{F(\map M\otimes \map I_\rR)~|~  \rR\in|\Transf|\}$.     Using this fact, we can identify the transformation $\map M$  with the linear map  
\begin{align}\label{super-m}
  G(\map M)  =  \bigoplus_{\rR\in  |\Transf|}      F  (  \map M\otimes \map I_\rR)  \, , 
\end{align}
which transforms elements of  the vector space  
\begin{align}\label{GA}
 G(\rA)   :  =   \bigoplus_{\rR\in  |\Transf|}        F(\rA\otimes \rR)      
\end{align}
into elements of the  vector space     
\begin{align}
 G(\rB)   :  =   \bigoplus_{\rR\in  |\Transf|}        F(\rB\otimes \rR)    \, .     
\end{align}
Endowing the vector spaces $G(\rA)$ and $G(\rB)$ with the direct sum cone inherited from the vector spaces $F(\rA\otimes \rR)$ and $F(\rB\otimes \rR)$, respectively, one has that $G(\map M)$ is a positive map.

Note that for two transformations $\map M  \in\Transf(\rA\to \rB)$ and $\map N \in  \Transf (\rB\to \rC)$ one has
\begin{align}\label{super-s}
G(\map N) \circ       G(\map M)    =    \bigoplus_{\rR\in  |\Transf|}      F  [( \map N\circ \map M)\otimes \map I_\rR]   \equiv   G(\map N\circ\map M) \,  ,
\end{align}
which shows that $G$ is a functor.   By definition \ref{super-m}, the functor $G$ is faithful.

In addition,  for two transformations    $\map A  \in\Transf(\rA\to \rA')$ and $\map B \in  \Transf (\rB\to \rB')$ one has   
\begin{align}\label{super-p}
G(\map A)  \otimes    G(\map B)     \simeq    \bigoplus_{\rR\in  |\Transf|}      F  (\map A \otimes \map B\otimes \map I_\rR)   \equiv   G (\map A\otimes \map B) \,  ,
\end{align}
up to natural isomorphism.    In other words, $G$ is a strong monoidal functor.     Summarizing, we obtained the following
\begin{theorem}\label{theo2}
Let $ \Transf$ be the category of transformations in a quotient theory.  Then, there exists a  faithful strong symmetric monoidal functor  $G$ between $\Transf$ and an SMC  of  partially order real vector spaces and positive linear maps, defined as in Eq. (\ref{super-m}).      The functor $G$  is such that, for every system $\rA$, the following conditions are satisfied:  
\begin{align}
    \Span_{\Reals}    \left\{ G  (  \alpha)   ~,~  \alpha \in \St (\rA)   \right\}    \simeq F(\rA)    \, ,
    \end{align} 
     \begin{align}
       \Span_{\Reals}    \left\{ G  (  a)   ~,~  a \in \Eff (\rA)   \right\}     \simeq  F(\rA)^*    \, ,
      \end{align}
and
\begin{align}\label{probG}  
 G (  a   \circ \rho)   & =        \Prob ( a\circ \rho)  ~   G(  {\sf 1})   \qquad \forall \alpha  \in  \St(\rA) \, ,   \forall  a\in\Eff  (\rA) \, .
%\nonumber   \Span_\Reals   \left\{      G(   \alpha)   , ~  \alpha \in \St (\rA)    \right\}    & \simeq       \Span_\Reals   \left\{      F(   \alpha)   , ~  \alpha \in \St (\rA)    \right\}  \\
 % \Span_\Reals   \left\{      G(   a)   , ~  a  \in \Eff (\rA)    \right\}   &   \simeq    \Span_\Reals   \left\{      F(   a)   , ~  a \in \Eff (\rA)    \right\}   \, .
%  \label{basic}
\end{align}
\end{theorem}

In other words, the category $\Transf$ can be \emph{identified} with an SMC of ordered vector spaces and positive maps.  
  This fact is very useful because it allows one to define  linear combinations of transformations, such as 
  \begin{align}
  \map X  =   \sum_x   \, x_i  \,     \map M_i  \qquad  \{x_i\}  \subset \Reals  \, ,   \{\map M_i\}  \subset   \Transf(\rA\to \rB) \, .
  \end{align}
 The real vector space spanned by $\Transf(\rA\to \rB)$ will be denoted $\Transf_\Reals  (\rA\to \rB)$.  
Note that, typically,    $\Transf_\Reals  (\rA\to \rB)$  is  a proper subspace of  $ L[ G(\rA),  G(\rB)]$, the space of all linear maps from the vector space $G(\rA)$ to the vector space $G(\rB)$.    For example, one has $ \Transf_\Reals  (\rI\to \rI)  \simeq \Reals$ whereas $L[G(\rI),  G(\rI)]$ is generally an infinite-dimensional vector space.    In the case of states and effects, we  use the special notations   $$\St_\Reals  (  \rA )  :  =   \Transf_\Reals  (\rI\to \rA) $$
and   $$  \Eff_\Reals  (\rA)  :  =  \Transf_\Reals(\rA\to \rI) \, .$$

We say that a system is \emph{finite-dimensional} iff the dimension of $\St_\Reals (\rA)$  [or, equivalently, the dimension of $\Eff_\Reals  (\rA) \,$] is finite.  

\subsection{Physicalizing the readout}  

In the basic framework we introduced outcome spaces  as abstract index sets labelling different possible processes.    At the intuitive level, however, the outcome of a test is written on some physical system and  can be read out from it.  In order to express this fact one can require that every test arises from a transformation followed by a measurement on one system. This requirement is expressed by the following 
\begin{ax}[Physicalization of readout]\label{ax:read}  
For every pair of systems $\rA,\rB\in|\Transf|$, every outcome space $ \rX\in\Outcomes$, and every test $\{  \map M_x \}_{x\in\rX}  \in  \Tests (\rA\to \rB,\rX)$ there exist a system $\rC  \in  |\Transf|$,  a transformation $\map M  \in \Transf(\rA\to \rB\otimes \rC)$, and a  test $\{c_x\}_{x\in\rX}  \in   \Tests (\rC\to \rI,\rX)$ such that  
\begin{equation}\label{read}
  \begin{aligned}  \Qcircuit @C=.5em @R=0em @!R 
{        &  \qw  \poloFantasmaCn{\rA}&     \gate{\map M_x }    &  \qw  \poloFantasmaCn{\rB}  &\qw  } 
\quad=  \quad 
 \Qcircuit @C=.5em @R=0em @!R {      &  \qw  \poloFantasmaCn{\rA}& \multigate{2}{\map M}    & \qw \poloFantasmaCn{\rB}  &  \qw    &&&&&&&&  \qquad \forall x\in\rX  \, .     \\
 & & \pureghost{\map M}   && 
  \\
& & \pureghost{\map M}    & \qw \poloFantasmaCn{\rC}  &  \measureD{c_x}} 
\end{aligned}     
\end{equation}
\end{ax}  
\medskip 

In principle, one could also require that  the test  $\{c_x\}_{x\in\rX}  \in   \Tests (\rC\to \rI,\rX)$ can distinguish perfectly among a set of states, as it was done in Ref. \cite{chiribella10}.   Similarly, one could set up more requirements on the transformation $\map M$,~e.~g.~requiring it to be of the form $\map M   =  \sum_{x\in\rX}    \,     \map M_x  \otimes \gamma_x$, where $ \{\gamma_x\}_{x\in\rX}$ is a set of perfectly distinguishable states.    However, these extra requirements are less  essential than the basic assumption  that the theory should be able to model the readout process, as  in Eq. (\ref{read}).

\section{Pure states, pure transformations,  and reversible transformations}

The framework of operational-probabilistic theories allows one to define the familiar notion of pure state:    a state  $\alpha\in\St(\rA)$ is \emph{pure} iff for every set of states $\{\alpha_i\}_{i=1}^N  \subset \St(\rA)$ one has the implication 
\begin{align}
  \sum_{i=1}^N   \alpha_i    =  \alpha        \quad \Longrightarrow  \quad      \alpha_i      =  p_i  \, \alpha   \, ,  \quad \forall  i\in\{1,\dots, N\}  \, ,       
\end{align}
for some probabilities $\{p_i\}$. 
% In words, a state is pure if it cannot be simulated by randomizing the preparation of the system. 
We denote the set of pure states of system $\rA$ as $\Pur\St(\rA)$.

A similar notion  can be put forward for transformations:  a transformation $\map M  \in\Transf(\rA\to \rB)$ is \emph{pure} iff for every set of transformations $\{  \map M_i\}_{i=1}^N \subset \Transf(\rA\to \rB)$ one has the implication
\begin{align}
  \sum_{i=1}^N   \map M_i    =  \map M       \quad \Longrightarrow  \quad      \map M_i      =  p_i  \, \map M \, ,  \quad \forall  i\in\{1,\dots, N\}\, ,       
\end{align}
for some probabilities $\{p_i\}$. We denote the set of pure transformations from $\rA$ to $\rB$ as $\Pur\Transf(\rA\to \rB)$. For effects, we use the notation $\Pur\Eff (\rA)   :  =  \Pur\Transf(\rA\to \rI)$.

Finally, another important notion is the notion of reversible transformation.  This is a primitive notion, which does not even need the probabilistic structure:  a reversible transformation is just an isomorphism in the category $\Transf$. Explicitly, a transformation $\map U  \in\Transf(\rA\to \rB)$ is \emph{reversible} iff  there exists another transformation $\map U^{-1}  \in \Transf(\rB\to \rA)$ such that 
\begin{align}
\begin{aligned}  
\Qcircuit @C=.5em @R=0em @!R {  & \qw \poloFantasmaCn{\rA}  &\gate{\map U}   &   \qw \poloFantasmaCn{\rB}   & \gate{ \map U^{-1}  }  &    \qw  \poloFantasmaCn{\rA}    &  \qw   } 
\end{aligned}   
=  
\begin{aligned}  
\Qcircuit @C=.5em @R=0em @!R {     & \qw \poloFantasmaCn{\rA}  &\gate{\map I}  &   \qw \poloFantasmaCn{\rA}   &  \qw  } 
\end{aligned}     
\qquad  {\rm and}  \qquad \begin{aligned}  
\Qcircuit @C=.5em @R=0em @!R {  & \qw \poloFantasmaCn{\rB}  &\gate{\map U^{-1}}   &   \qw \poloFantasmaCn{\rA}   & \gate{ \map U  }  &    \qw  \poloFantasmaCn{\rB}    &  \qw   } 
\end{aligned}   
=  
\begin{aligned}  
\Qcircuit @C=.5em @R=0em @!R {     & \qw \poloFantasmaCn{\rB}  &\gate{\map I}  &   \qw \poloFantasmaCn{\rB}   &  \qw  } 
\end{aligned}    \quad .  
\end{align}   
The set of reversible transformations of type $\rA\to \rB$ will be denoted by $\Rev\Transf(\rA\to \rB)$.
The notion of pure state, pure transformation, and reversible transformation will be used  in section \ref{sec:puri} for  the formulation of the Purification Principle. 
   
 \section{Causality}  
 
 Once the basic framework of operational-probabilistic theories has been defined, one can  formulate axioms that define classes of theories sharing common features.   
 A particularly basic axiom is causality, which reads as follows 
 \begin{ax}[Causality]\label{ax:causality}
For every system $\rA\in |\Transf| $  there exists an effect   $\Tr_\rA \in\Eff(\rA)$, called the \emph{trace} (or the \emph{deterministic effect} \cite{chiribella10,chiribella11})  and graphically represented as   $\Qcircuit @C=.5em @R=0em @!R {     &   \qw \poloFantasmaCn{\rB}   &  \measureD {\Tr} }$  ,  such that 
\begin{align}\label{norm}
  \sum_{x\in\rX}   a_x   =      \Tr_\rA  \qquad \qquad \forall \{a_x\}_{x\in\rX}   \in   \Tests  (\rA\to \rI ,  \rX)\, ,  \qquad  \forall \rX\in\Outcomes    \, .      
\end{align}
 \end{ax}
From the definition and from the fact that $\Tests$ is an SMC,  if follows immediately that the trace satisfies the following relations  
\begin{align}
\nonumber  \Tr_{\rA\otimes \rB}    & =    \Tr_\rA  \otimes \Tr_{\rB}  \\
\Tr_\rI   &  =     {\sf 1} \, .\label{trace}
\end{align}

\medskip  

\subsection{Deterministic transformations}   We call a transformation  $\map M   \in\Transf(\rA\to \rB)$ \emph{deterministic}\footnote{In the original works \cite{chiribella10,chiribella11}  the deterministic transformations were defined in a different way, starting from tests with singleton outcome set.   However, in a quotient theory where the causality axiom holds, the definition of Refs. \cite{chiribella10,chiribella11} is equivalent to the one given here. } iff  \begin{align}\label{normalization}
\begin{aligned}  
\Qcircuit @C=.5em @R=0em @!R {     & \qw \poloFantasmaCn{\rA}  &\gate{\map M}   &   \qw \poloFantasmaCn{\rB}   & \measureD {\Tr}} 
\end{aligned}   
&= 
\begin{aligned}  
\Qcircuit @C=.5em @R=0em @!R {     & \qw \poloFantasmaCn{\rA}  & \measureD {\Tr}} 
\end{aligned}    \, .  
\end{align}  
Combining the definition with Eqs. (\ref{trace})  one obtains that the deterministic transformations form a symmetric monoidal subcategory of $\Transf$, which we denote by $\Det\Transf$.       An equivalent way to state the causality axiom, put forward by Coecke and Lal in Refs. \cite{lal,coecke},  is to start from a distinguished SMC   $\Det\Transf  \hookrightarrow   \Transf$ and to impose  that the tensor unit $\rI$ is terminal in   $\Det\Transf $.    
 
 \medskip 
 
\subsection{Marginal states and marginal transformations}  

The trace allows one to introduce a canonical notion of marginal state:  given a state $\sigma\in\St (\rA\otimes \rB)$, the marginal of \emph{$\sigma$ on system $\rA$} is the state $\rho\in\St(\rA)$  defined by  
\begin{equation*}
\begin{aligned}  \Qcircuit @C=.5em @R=0em @!R 
{ & \prepareC{\rho }    &  \qw  \poloFantasmaCn{\rA}  &\qw  } 
\end{aligned} 
:=
\begin{aligned}  \Qcircuit @C=.5em @R=0em @!R { & \multiprepareC{2}{\sigma}    & \qw \poloFantasmaCn{\rA}  &  \qw \\
& \pureghost{\sigma}   &&\\
 & \pureghost{\sigma}    & \qw \poloFantasmaCn{\rB}  &  \measureD{\Tr}} 
\end{aligned}   \quad .   
\end{equation*}
 Conversely, we say that $\sigma$ is an \emph{extension of $\rho$ to system $\rA\otimes \rB$.}  
 
 The same definition can be phrased for general transformations: the \emph{marginal of a transformation $\map N  \in \Transf(\rA\to \rB\otimes \rC)$   on system $\rB$} is the transformation $\map M\in\Transf(\rA\to \rB)$  defined  as 
\begin{equation}
\begin{aligned}  \Qcircuit @C=.5em @R=0em @!R 
{    &\qw  \poloFantasmaCn{\rA}  &\gate{\map M}   &    \qw  \poloFantasmaCn{\rB}  &  \qw  } 
\end{aligned} 
:=
\begin{aligned}  \Qcircuit @C=.5em @R=0em @!R {    
&\qw  \poloFantasmaCn{\rA} & \multigate{2}{\map N}    & \qw \poloFantasmaCn{\rB}  &  \qw \\
&& \pureghost{\map N}   &&&\\
&& \pureghost{\map N}    &   \qw \poloFantasmaCn{\rC}  &  \measureD{\Tr}} 
\end{aligned}     \quad .
\end{equation}
 When this is the case, we say that $\map N$ is an extension of   $\map M$ to system $\rB\otimes \rC$.

\section{Purification}\label{sec:puri}

Purification is the requirement that every mixed state can be modelled as the marginal of a pure state in an essentially unique way.  For finite-dimensional systems, the axiom reads: 

\begin{ax}[Purification]\label{ax:puri}
For every system $\rA\in  |\Transf|$ and for every state $\rho  \in  \St(\rA)$ there exists a system $\rB$, called the \emph{purifying system}, and a pure state $\Psi  \in\Pur\St(\rA\otimes \rB)$ such that 
\begin{equation*}
\begin{aligned}  \Qcircuit @C=.5em @R=0em @!R { & \multiprepareC{2}{\Psi}    & \qw \poloFantasmaCn{\rA}  &  \qw \\
& \pureghost{\Psi}   &&\\
 & \pureghost{\Psi}    & \qw \poloFantasmaCn{\rB}  &  \measureD{\Tr}} 
\end{aligned}    =  \begin{aligned}  \Qcircuit @C=.5em @R=0em @!R 
{ & \prepareC{\rho }    &  \qw  \poloFantasmaCn{\rA}  &\qw  } 
\end{aligned}   \quad .   
\end{equation*}
Moreover,  for every  system $\rB\in|  \Transf|$ and for every pair of pure states $\Psi,\Psi'\in\Pur\St(\rA\otimes \rB)$  one has the implication 
\begin{equation*}
\begin{aligned}  \Qcircuit @C=.5em @R=0em @!R { & \multiprepareC{2}{\Psi'}    & \qw \poloFantasmaCn{\rA}  &  \qw \\
& \pureghost{\Psi'}   &&\\
 & \pureghost{\Psi'}    & \qw \poloFantasmaCn{\rB}  &  \measureD{\Tr}} 
\end{aligned}   
=  \begin{aligned}  \Qcircuit @C=.5em @R=0em @!R { & \multiprepareC{2}{\Psi}    & \qw \poloFantasmaCn{\rA}  &  \qw \\
& \pureghost{\Psi}   &&\\
 & \pureghost{\Psi}    & \qw \poloFantasmaCn{\rB}  &  \measureD{\Tr}} 
\end{aligned}  \qquad \Longrightarrow \qquad   
\begin{aligned}  \Qcircuit @C=.5em @R=0em @!R { & \multiprepareC{2}{\Psi'}    & \qw \poloFantasmaCn{\rA}  &  \qw \\
& \pureghost{\Psi'}   &&\\
 & \pureghost{\Psi'}    & \qw \poloFantasmaCn{\rB}  &  \qw} 
\end{aligned}    = 
\begin{aligned}  \Qcircuit @C=.5em @R=0em @!R { & \multiprepareC{2}{\Psi}    & \qw \poloFantasmaCn{\rA}  &  \qw   &\qw &\qw \\
& \pureghost{\Psi}   &&  &&  \\
 & \pureghost{\Psi}    & \qw \poloFantasmaCn{\rB}  &  \gate{\tU}   &    \qw\poloFantasmaCn{\rB}  &\qw  } 
\end{aligned}   
\end{equation*}
for a reversible transformation $\map U  \in  \Rev\Transf  (\rB)$. 
\end{ax}

The Purification Axiom is the operational translation of  (some features of) the GNS construction  \cite{GN,S} in the special case of  finite-dimensional irreducible matrix algebras. 

\subsection{Consequences of purification}

Given the importance of the GNS construction it is probably not surprising that the Purification Axiom has a large number of consequences, some of which are listed in the following.  
 
\subsubsection{Transitivity of reversible transformations on pure states}  
   An immediate consequence of purification is that every two pure states of a system are connected by a reversible transformation: 
   \medskip 
\begin{proposition}
Suppose that the theory $\Theta$ satisfies purification. Then, for every system $\rA\in |\Transf|$ and every pair of pure states $\alpha, \alpha'\in\Pur\St(\rA)$ there exists a reversible transformation $\map U\in\Rev\Transf(\rA)$ such that  $  \alpha'  =  \map U\circ  \alpha $.  
\end{proposition}
% In general, for a   $*$-algebra $\map A$ and for a state $\rho$ on the algebra, the GNS construction provides a Hilbert space $\spc H$,  a unit vector $|\Psi\>$, and a $*$-representation  $\pi$  such that   
%\begin{align}\label{puri}
%\rho  (   A)     =        \< \Psi  |    \,  \pi  (A)  \,    |\Psi\>  \qquad \forall A  \in  \map A 
%\end{align}  
%and 
%\begin{align*}
%\spc H  =   \Span_{\mathbb  C}  \{    \pi  (A)  |\Psi\>   ,  \,    A\in \map A  \} \, .
%\end{align*}  
%Consider the algebra    $\map A  =  M_d(\mathbb{C})$, consisting  of all $d\times d$ matrices over the complex numbers.   The states on the algebra are identified with density matrices,  i.~e.~ positive semidefinite matrices with unit trace.  
%For a given density matrix   $\rho\in M_d(\mathbb{C})$, the GNS construction reads $  \spc H  =   $

This property had been used as an axiom in  Hardy's 2001 axiomatization \cite{hardy2001} and in subsequent axiomatizations  \cite{dakic2011,masanes2011}. 

\medskip  
 
\subsubsection{Steering}    Another consequence of purification  the fact that every decomposition of a mixed state can be induced by a measurement on the purifying system:
\medskip 
     
\begin{proposition}\label{prop:steer}  
Suppose that the theory $\Theta$ satisfies purification. Let  $\rho  \in\St(\rA)$ be a state of a generic system $\rA$ and let $\Psi \in\Pur\St(\rA\otimes \rB)$ be a purification of $\rho$.   Then, for  every outcome space $ \rX\in\Outcomes$ and every test $\{  \rho_x \}_{x\in\rX}  \in  \Tests (\rA\to \rB,\rX)$ such that $\sum_{x\in\rX}  \rho_x  =  \rho$ there exists a  measurement $\{b_x\}_{x\in\rX}  \in   \Tests (\rB\to \rI,\rX)$ such that  
\begin{equation*}
  \begin{aligned}  \Qcircuit @C=.5em @R=0em @!R 
{        &     \prepareC{\rho_x }    &  \qw  \poloFantasmaCn{\rA}  &\qw  } 
\quad =  \quad   \Qcircuit @C=.5em @R=0em @!R {      & \multiprepareC{2}{\Psi}    & \qw \poloFantasmaCn{\rA}  &  \qw   &&&&&&&  \forall x\in\rX  \, .\\
  & \pureghost{\Psi}   &&\\
 & \pureghost{\Psi}    & \qw \poloFantasmaCn{\rB}  &  \measureD{b_x}} 
\end{aligned}   
\end{equation*}
\end{proposition}  

\section{Faithfulness}

Until now we did not assume the existence of  mixed states: the general framework of operational-probabilistic theories describes also theories that have no mixed states at all, such as deterministic classical computation.   As a matter of fact, even the purification axiom is satisfied by deterministic classical computation---in a trivial way, since there is no mixed state to be purified there.   

We now require that in our theory   there exist mixed states and, in particular, that there exist states that are ``sufficiently mixed", according to the following definition:  we call a state $\omega\in\St(\rA)$  \emph{faithful} iff for every system $\rB$ and for every pair of transformations $\map M,\map M'\in\Transf(\rA\to \rB)$  the condition 
\begin{align}
\begin{aligned}  \Qcircuit @C=.5em @R=0em @!R { & \multiprepareC{2}{\sigma}    & \qw \poloFantasmaCn{\rA}  &  \gate{\map M}  &  \qw \poloFantasmaCn{\rB}  &\qw \\
& \pureghost{\sigma}   &&&&\\
 & \pureghost{\sigma}    & \qw \poloFantasmaCn{\rC}  &  \qw &\qw&\qw&} 
\end{aligned}       
=  \begin{aligned}  \Qcircuit @C=.5em @R=0em @!R { & \multiprepareC{2}{\sigma}    & \qw \poloFantasmaCn{\rA}  &  \gate{\map M'}  &  \qw \poloFantasmaCn{\rB}  &\qw \\
& \pureghost{\sigma}   &&&&\\
 & \pureghost{\sigma}    & \qw \poloFantasmaCn{\rC}  &  \qw &\qw&\qw&} 
\end{aligned}     \quad \forall  \rC  \in   |\Transf|  \, , \forall  \sigma  \in  \St(\rA\otimes \rC)  \,  :       \sigma  ~{\rm is~an~extension~of~} \omega     
 \end{align}
implies  $\map M=  \map M'$.    The definition of faithful states given here coincides with the $C^*$-algebraic  definition faithful states in the finite dimensional setting.    Faithful states exist in every convex theory where the set of states $\St(\rA)$ is convex and finite-dimensional \cite{chiribella10,chiribella11}:  in that case, they are nothing but the states in the interior of the  state space.     Rather that requiring convexity, here we just require the existence of a faithful state for every system: 
\begin{ax}[Faithfulness]\label{ax:faith}
For every system $\rA\in  |\Transf|$, there exists at least one faithful state $\omega_\rA  \in \St(\rA)$.  
\end{ax}

\section{Consequences of purification and faithfulness} 
The full power of purification appears in conjunction with the faithfulness axiom.  The core result is the existence of a pure state that establishes an injective correspondence between transformations and states. 

\subsection{The state-transformation isomorphism}
Once the definitions are in place, it is easy to establish a general isomorphism that has the operational features of the Choi isomorphism for finite-dimensional matrix algebras \cite{choi1975}:  

\begin{proposition}
Suppose that the theory $\Theta$ satisfies Axioms \ref{ax:read}, \ref{ax:puri}, and \ref{ax:faith}.  Then, for every system $\rA\in|\Transf |$ and every  purification of the faithful state of $\rA$,  say $\Psi\in\Pur\St(\rA\otimes \rC)$, has the following property: 
\begin{align}
\begin{aligned}  \Qcircuit @C=.5em @R=0em @!R { & \multiprepareC{2}{\Phi}    & \qw \poloFantasmaCn{\rA}  &  \gate{\map M}  &  \qw \poloFantasmaCn{\rB}  &\qw \\
& \pureghost{\Phi}   &&&&\\
 & \pureghost{\Phi}    & \qw \poloFantasmaCn{\rC}  &  \qw &\qw&\qw&} 
\end{aligned}       
=  \begin{aligned}  \Qcircuit @C=.5em @R=0em @!R { & \multiprepareC{2}{\Phi}    & \qw \poloFantasmaCn{\rA}  &  \gate{\map M'}  &  \qw \poloFantasmaCn{\rB}  &\qw \\
& \pureghost{\Phi}   &&&&\\
 & \pureghost{\Phi}    & \qw \poloFantasmaCn{\rC}  &  \qw &\qw&\qw&} 
\end{aligned}    
\quad   \Longrightarrow   \quad     \map M  =  \map M'    \quad  \forall \rB  \in  |\Transf|    ,  \,  \forall  \map M,\map M' \in \Transf(\rA\to \rB)  \, .  
\end{align}   
\end{proposition}

We call the correspondence $ \map M   \mapsto      \Phi_{\map M} :  =    (  \map M\otimes \map I_\rC)  \circ  \Phi$ the    \emph{ state-transformation isomorphism} \cite{chiribella10,chiribella11}.   
The fact that the state-transformation isomorphism is set up by a \emph{pure} state  $\Phi$  has major consequences, briefly sketched in the next subsections:

\subsection{No information without disturbance}
 The most consequence of the    pure  state-transformation isomorphism  is the No Information Without Disturbance principle, which states that every test $\{\map M_x\}_{x\in\rX}\in\Tests(\rA\to \rA,\rX)$ satisfying  the ``no disturbance condition"
\begin{align*}
\sum_{x\in\rX}\map M_x  =  \map I_\rA   
\end{align*} 
must also satisfy the ``no information condition"  
\begin{align*}
\map M_x   =  p_x  \,  \map I_\rA  \qquad \forall x\in\rX  
\end{align*}
for some probability distribution $\{p_x\}_{x\in\rX}$.

\subsection{Purification of transformations}

We have seen that the purification of states expresses the operational content of the GNS construction for finite dimensional quantum systems.   What about Stinespring's theorem \cite{stine}, whose statement includes the GNS construction as a special case?     Interestingly,  using the state-transformation isomorphism in combination with axioms \ref{ax:read} and \ref{ax:faith}   one can obtain the operational version of Stinespring's theorem from the purification axiom:  

\begin{proposition}[Purification of transformations]\label{ax:puritransf}
Suppose that the theory $\Theta$ satisfies Axioms \ref{ax:read}, \ref{ax:puri}, and \ref{ax:faith}.  
For every pair of systems $\rA,\rB\in  |\Transf|$ and for every transformation $\map M  \in  \Transf(\rA\to \rB)$ there exists a system $\rC$ and a pure transformation $\map P  \in\Pur\Transf(\rA  \to \rB\otimes \rC)$ such that 
\begin{equation*}
\begin{aligned}  \Qcircuit @C=.5em @R=0em @!R {      &  \qw  \poloFantasmaCn{\rA}& \multigate{2}{\map P}    & \qw \poloFantasmaCn{\rB}  &  \qw \\
 & & \pureghost{\map P}   &&\\
& & \pureghost{\map P}    & \qw \poloFantasmaCn{\rC}  &  \measureD{\Tr}} 
\end{aligned}    =  \begin{aligned}  \Qcircuit @C=.5em @R=0em @!R 
{        &  \qw  \poloFantasmaCn{\rA}&     \gate{\map M }    &  \qw  \poloFantasmaCn{\rB}  &\qw  } 
\end{aligned}   \quad .   
\end{equation*}
Moreover,  for every  system $\rC\in|  \Transf|$ and for every pair of pure transformations  $\map P, \map P'  \in\Pur\Transf(\rA\otimes \rB)$  one has the implication 
\begin{equation*}
\begin{aligned}  \Qcircuit @C=.5em @R=0em @!R {        &  \qw  \poloFantasmaCn{\rA}&     \multigate{2}{\map P'}    & \qw \poloFantasmaCn{\rB}  &  \qw \\
&& \pureghost{\map P'}   &&\\
      & &    \pureghost{\map P'}    & \qw \poloFantasmaCn{\rC}  &  \measureD{\Tr}} 
\end{aligned}   
=  \begin{aligned}  \Qcircuit @C=.5em @R=0em @!R {       &  \qw  \poloFantasmaCn{\rA}& \multigate{2}{\map P}    & \qw \poloFantasmaCn{\rA}  &  \qw \\
&& \pureghost{\map P}   &&\\
       &     & \pureghost{\map P}    & \qw \poloFantasmaCn{\rC}  &  \measureD{\Tr}} 
\end{aligned}  \qquad \Longrightarrow \qquad   
\begin{aligned}  \Qcircuit @C=.5em @R=0em @!R {         &  \qw  \poloFantasmaCn{\rA}&  \multigate{2}{\map P'}    & \qw \poloFantasmaCn{\rB}  &  \qw \\
&& \pureghost{\map P'}   &&\\
      &  &  \pureghost{\map P'}    & \qw \poloFantasmaCn{\rC}  &  \qw} 
\end{aligned}    = 
\begin{aligned}  \Qcircuit @C=.5em @R=0em @!R {      &  \qw  \poloFantasmaCn{\rA} & \multigate{2}{\map P}    & \qw \poloFantasmaCn{\rB}  &  \qw   &\qw &\qw \\
&& \pureghost{\map P}   &&  &&  \\
      &   & \pureghost{\map P}    & \qw \poloFantasmaCn{\rC}  &  \gate{\tU}   &    \qw\poloFantasmaCn{\rC}  &\qw  } 
\end{aligned}   
\end{equation*}
for a reversible transformation $\map U  \in  \Rev\Transf  (\rC)$. 
\end{proposition}

\medskip  

This result concludes our review of dilation-type results that can derived axiomatically in the framework of operational-probabilistic theories.  As the reader might have noticed, we did not discuss dilation theorems of the Naimark-  \cite{naimark} and Ozawa-type \cite{ozawa}.  They are based on a different conceptual ingredient, namely the notion of \emph{sharp measurement}.  A discussion in this direction can be found in Ref.\cite{sharp}, which derived  upper bounds on quantum nonlocality and contextuality  from a Naimark-type dilation of measurements to sharp measurements. 

\medskip

{\bf Acknowledgements.}    This work was supported   by the National Basic Research Program of China
(973) 2011CBA00300 (2011CBA00301) and  by the National
Natural Science Foundation of China (Grants 11350110207,
61033001, 61061130540), by the 1000 Youth Fellowship
Program of China, by the Foundational Questions Institute (Grants FQXi-RFP3-1325). GC acknowledges the hospitality of Perimeter Institute for Theoretical Physics.  Research at Perimeter Institute for Theoretical Physics is supported in part by the Government of Canada through NSERC and by the Province of Ontario through MRI.

\nocite{*}
\bibliographystyle{eptcs}
%\bibliography{generic}

\end{document}